\documentclass[preprintnumbers,amsmath,11pt,onecolumn, amssymb,floatfix,superscriptaddress,nofootinbib]{revtex4}
\usepackage{graphicx}
\usepackage{epsfig}
\usepackage{bm}
\usepackage{amsfonts}
\def\be {\begin{equation}}
\def\ee {\end{equation}}
\def\ba {\begin{eqnarray}}
\def\ea {\end{eqnarray}}

\begin{document}
\title{Thermodynamics of a Schwarzschild Black Hole in Phantom Cosmology with Entropy Corrections}

 \author{\textbf{Mubasher Jamil}}\email{mjamil@camp.nust.edu.pk}
\affiliation{Center for Advanced Mathematics and Physics (CAMP),
National University of Sciences and Technology (NUST), H-12,
Islamabad, Pakistan}\affiliation{Eurasian International Center
for Theoretical Physics, Eurasian National University, Astana
010008, Kazakhstan}

 \author{\textbf{D. Momeni}}\email{d.momeni@yahoo.com}\affiliation{Eurasian International Center
for Theoretical Physics, Eurasian National University, Astana
010008, Kazakhstan}
\author{\textbf{Kazuharu Bamba}}
\email{bamba@kmi.nagoya-u.ac.jp}
\affiliation{Kobayashi-Maskawa Institute for the Origin of Particles and the Universe, Nagoya
University,
Nagoya 464-8602, Japan}
 \author{\textbf{Ratbay Myrzakulov}}\email{
rmyrzakulov@csufresno.edu}\affiliation{Eurasian International Center
for Theoretical Physics, Eurasian National University, Astana
010008, Kazakhstan}

\begin{abstract}
\textbf{Abstract:} Motivated by some earlier works
\cite{pavon,sadjadi}  dealing with the study of generalized second
law (GSL) of thermodynamics for a system  comprising of a
Schwarzschild black accreting a test non-self-gravitating fluid
namely phantom energy in FRW universe, we extend them when the
entropy of horizons of black hole and the cosmological undergo
quantum corrections. Two types of such corrections are relevant here
including logarithmic and power-law, while both are motivated from
different theoretical backgrounds. We obtain general mathematical
conditions for the validity of GSL in each case. Further we find
that GSL restricts the mass of black hole for accretion of phantom
energy. As such we obtain upper bounds on the mass of black hole
above which the black hole cannot accrete the phantom fluid,
otherwise the GSL is violated.
\end{abstract}

\maketitle
\newpage

\section{Introduction}

Modern approaches to unify theories of quantum mechanics and general
relativity,  for instance, string theory and loop quantum gravity
predict that a black hole emits thermal radiations whose thermal
spectrum might deviate from
  Planck black body spectrum at Planck scale \cite{barton}.
   The black hole's event horizon possesses temperature inversely proportional to black hole mass and with an entropy proportional to its horizon's surface area (in units $c=G=\hbar=1$) i.e.
 \begin{equation}\label{f61}
S_h=\frac{A_h}{4},
\end{equation}
where $A_h=4\pi R_h^2$ is the area of the black hole's event
horizon.  Therefore the horizon entropy (\ref{f61}) becomes
\begin{equation}\label{f6}
S_h=\pi R_h^2.
\end{equation}
 These seminal connections between black holes and thermodynamics were
   initially made by  Hawking and Bekenstein several decades ago \cite{hawking}.
    The Hawking temperature and horizon entropy together with the black
hole mass obey the first law of thermodynamics $TdS=dE+pdV$.
Padmanabhan showed that Einstein field equations for a spherically
symmetric spacetime can be recast in the form of first law of
thermodynamics \cite{paddy}. Cai \& Kim applied the similar
formalism and demonstrated that by applying the first law of
thermodynamics to the apparent horizon of a
Friedmann-Robertson-Walker universe the Friedmann equations of the
universe with any spatial curvature can be derived from the first
law \cite{cai}. Recently similar results have been obtained in
scalar tensor, Gauss-Bonnet, Lovelock and $f(R)$ gravities by
different authors \cite{akbar1}. Jacobson showed that Einstein field
equation is nothing but an equation of state of spacetime i.e. the
Einstein equation can be derived by assuming the universality of
(\ref{f61}) on any local Rindler horizons \cite{ted}. In black hole
physics, the generalized second law is a conjecture about black hole
thermodynamics which states that ``the sum of the black hole entropy
(1/4 of the horizon area) and the common (ordinary) entropy in the
black hole exterior never decreases' originally proposed by
Bekenstein \cite{hod,b}. However as discussed by Jacobson
\cite{ted}, the entire framework of black hole thermodynamics and,
in particular, the notion of black hole entropy extends to any
causal horizon. In cosmological spacetime, the corresponding object
is \textit{apparent horizon} \cite{peng}.

The power-law correction to entropy which appear in
dealing with the entanglement of quantum fields in and
out the horizon is given by
is \cite{power}
\begin{equation}\label{1b}
S_h=\frac{A_h}{4}\Big(1-K_\alpha A_h^{1-\frac{\alpha}{2}} \Big),
\end{equation}
where $\alpha$ is a dimensionless constant and $A_h=4\pi R_h^2$ is the area while $R_h$ is the radius of the horizon.
\begin{equation}\label{1c}
K_\alpha=\frac{\alpha(4\pi)^{\frac{\alpha}{2}-1}}{(4-\alpha)r_c^{2-\alpha}}.
\end{equation}
where $r_c$ is a cross-over scale, $R_h$ is the radius and $A_h$ is
area of the cosmological horizon. For entropy to be a well-defined
quantity, we require $\alpha>0$.  The second term in (\ref{1b}) can
be regarded as a power-law correction to the area law, resulting
from entanglement, when the wave-function of the field is chosen to
be a superposition of ground state and exited state \cite{p1}.
Several aspects of power-law corrected entropy (\ref{1b}) have been
studied in literature including GSL \cite{jamil3}, power-law entropy
corrected models of dark energy \cite{jamil4}.

The quantum corrections provided to the entropy-area relationship
leads to  the curvature correction in the Einstein-Hilbert action
and vice versa. The logarithmic corrected entropy is  \cite{jamil5}
\begin{equation}\label{1a}
S_h=\frac{A_h}{4}+\beta\log\Big( \frac{A_h}{4} \Big)+\gamma.
\end{equation}
These corrections arise in the black hole entropy in loop quantum
gravity  due to thermal equilibrium fluctuations and quantum
fluctuations. Jamil \& Sadjadi \cite{log} showed that in a (super)
accelerated universe GSL is valid whenever $\beta(<)>0$ leading to a
(negative) positive contribution from logarithmic correction to the
entropy. In the case of super acceleration the temperature of the
dark energy is obtained to be less or equal to the Hawking
temperature. Using the corrected entropy-area relation motivated by
the loop quantum gravity, Karami et al \cite{karami} investigated
the validity of the GSL in the FRW universe filled with an
interacting viscous dark energy with dark matter and radiation. They
showed that GSL is always satisfied throughout the history of the
universe for any spatial curvature regardless of the dark energy
model.

We consider a scenario of a spatially flat, homogeneous, isotropic
universe filled with phantom energy and contain a Schwarzschild
black hole. Since this simple cosmic system consists of three
components, we associate the entropy with each component. The
entropy of Schwarzschild black hole and FRW universe is proportional
to the size (area) of their horizon (only if the entropic
corrections are ignored) while for phantom energy, the entropy is
calculated via the first law of thermodynamics. We assume that the
black hole accretes phantom energy such that the mass of black hole
decreases very slowly while preserving the spherical symmetry. This
kind of accretion is termed as quasi-static accretion, and the
corresponding black hole in quasi-static state (i.e. Schwarzschild
geometry is still valid) \cite{horvath}. While accretion, the
entropies of phantom energy and the black hole vary, but the total
entropy of the system remains non-decreasing. From the argument of
first law of thermodynamics, we notice that the rate of change of
entropy of phantom energy $\dot S_d=T^{-1}\dot H R_h^2$ depends on
its temperature $T$ and the rate of change of Hubble parameter $\dot
H$. The entropy $\dot S_d>0$ if both $T>0(<0)$ and $\dot H>0(<0)$.
According to some earlier thermodynamic approaches to gravity, the
form of entropy-area relation applies the same to different
horizons, here we apply the same principle in different sections.
First we choose the classical Bekenstein-Hawking relation for
entropy for the horizons of both black hole and cosmological and
investigate the GSL. Later we study the same phenomenon by including
the power-law and logarithmic corrections to the entropy of both
black hole and cosmological horizon. Although the horizon of a
Schwarzschild black hole is uniquely defined, the form of
cosmological horizon is not so. We use the two well-known forms of
cosmic horizons i.e. the future event horizon and the apparent
horizon (both will be defined later). 
The idea of the 
combined effect of a cosmological system involving a dark energy
component and a black hole has also been explored in `entropic
cosmology' \cite{yifu}. In these works, Cai and collaborators
investigated the dynamical thermal balance of a double-screen model
(corresponding to cosmic horizon and black hole horizon) which can
realize both inflation and late time acceleration of the Universe.

Earlier Izquierdo \& Pavon \cite{pavon} investigated the GSL for a
system  comprising of a Schwarzschild black accreting phantom energy
in FRW universe. They showed that GSL is violated. Later Sadjadi
\cite{sadjadi} investigated the same problem and showed that GSL
will be satisfied if the temperature is not taken as de Sitter
temperature. Understanding the evolution of a black hole in a FRW
cosmological background is a very old problem starting from Hawking
\& Carr \cite{carr} whose satisfactory resolution is still not
available, however several approximations (like the present
analysis) are available in the literature, e.g.
\cite{sadjadi,pavon,jamil}.

The plan of our paper is as follows: In section-II, we write down
the  basic equations of standard model of cosmology and the
definition of generalized second law of thermodynamics in the
present context. In sections-III, IV and V, we study the GSL with
Bekenstein-Hawking entropy-area relation, power-law entropy
correction and logarithmic entropic correction, respectively with
the use of apparent and event horizons. In section-VI, we discuss
the constraints imposed by GSL on the black hole mass for accretion
of phantom energy. In last section, we write down the conclusion
giving a summary of our results.

\section{Basic Equations}

Assuming homogeneous, isotropic and spatially flat Friedmann-Robertson-Walker metric:
\begin{equation}\label{f1}
ds^2=-dt^2+a(t)^2(dr^2+r^2(d\theta^2+\sin^2\theta d\phi^2)).
\end{equation}
The Friedmann equations are
\begin{equation}\label{f2}
H^2=\frac{8\pi }{3}\rho,
\end{equation}
\begin{equation}\label{f3}
\dot H=-4\pi (\rho+p).
\end{equation}
The continuity equation is
\begin{equation}\label{f4}
\dot\rho+3H(\rho+p)=0,
\end{equation}
where $\rho$ and $p$ are the energy density and pressure of phantom
energy.  Assuming the phantom energy as a perfect fluid, we specify
it by a phenomenological equation of state
\begin{equation}\label{f5}
p=w\rho,
\end{equation}
where $w<-1$ is the dimensionless state parameter of phantom energy.
The notion of phantom energy was introduced by Caldwell et al
\cite{cald} as a separate candidate of explaining cosmic
acceleration. The phantom energy possesses some esoteric properties:
``Phantom energy rips apart the Milky Way, solar system, Earth, and
ultimately the molecules, atoms, nuclei, and nucleons of which we
are composed, before the death of the Universe in a Big Rip''
\cite{cald}. Although the model of phantom energy is consistent with
the observational data \cite{cald1}, the cosmic doomsday (or Big
Rip) can be avoided in certain theories of modified gravity
\cite{odintsov}. Thermodynamical studies show that phantom energy
possesses negative temperature and positive entropy \cite{sigu}.
However some discussion contrary to \cite{sigu} on phantom
thermodynamics has been performed in \cite{lima1}. Nojiri et al
discussed the occurrence of different types of future singularities
in phantom cosmology \cite{noji}. One of these singularities is Big
Rip described as: (or Type-I): As $t\rightarrow t_s$,
$a\rightarrow\infty$, $\rho\rightarrow\infty$,
$|p|\rightarrow\infty$. A form of scale factor satisfying these
conditions can be $a(t)=a_0(t_s-t)^{n}$, $t_s>t$,
$n=\frac{2}{3(1+w)}<0$, $a_0>0$ where $t_s$ is the Big Rip time. The
corresponding Hubble parameter goes like
\begin{equation}\label{h}
H(t)=\frac{2}{3(1+w)(t-t_s)}. \ \ (w<-1)
\end{equation}
Thus Hubble parameter diverges as $t\rightarrow t_s$.

Babichev et al \cite{babi} demonstrated that the accretion of
phantom energy  as a test perfect fluid on a stationary spherically
symmetric black hole gradually reduces the mass of black hole.
Further the mass of black holes approach to zero near the time
called Big Rip. They deduced that the rate of change in black hole
mass due to phantom energy accretion goes like \cite{babi}
\begin{equation}\label{f13}
\dot M=4\pi A r_h^2(\rho+p)<0.
\end{equation}
Above $A$ is a positive dimensional constant, $r_h$ is the
Schwarzschild radius, while energy density and pressure of phantom
energy violates the null energy condition $\rho+p<0$. Later on their
analysis extended in various ways: using bulk viscosity, generalized
Chaplygin gas, Riessner-Nordstrom and Kerr-Newmann black holes
\cite{jamilbh} to list a few.

Making use of (\ref{f3}) in (\ref{f13}), we can write
\begin{equation}
\dot M=-4AM^2\dot H.
\end{equation}
On integration, we obtain
\begin{equation}\label{f14}
M(H)=\frac{1}{C_1+4AH},
\end{equation}
where $C_1$ is a constant of integration. From (\ref{f14}),  we
observe that mass of black hole decreases as the rate of cosmic
expansion increases.

For generalization of these results to many black holes, one
can follow the procedure of Khan and Israel \cite{khan} by first
replacing the spherically symmetric black hole by a point mass
located on the $z$-axis. Then we can use the superposition principle
to write the general expression of gravitational potential of the
system. In this case, we can replace the mass of a BH by the total
mass of many point particles. However we are not interested to such
extensions and beyond the scope of our paper. 

To calculate the entropy of phantom fluid, we use the first  law of
thermodynamics which relates the pressure, energy density, total
energy and temperature of phantom energy i.e.
\begin{equation}\label{f16a}
dS_d=\frac{1}{T}(dE+pdV)=\frac{1}{T}((\rho+p)dV+Vd\rho).
\end{equation}
The form of GSL containing the time derivatives of entropies of
black hole's  horizon, phantom energy and cosmological's horizon is
\begin{equation}\label{f16}
\dot S_{tot}\equiv\dot S_{BH}+\dot S_{d}+\dot S_{A}\geq0.
\end{equation}
Above $\dot S_{tot}$ is the rate of change of total entropy which
must be non-decreasing. 
Here we would like to comment that
as a result of accretion, the dark energy goes inside the BH event
horizon. Since the major bulk of dark energy density lies outside
the BH horizon than to its interior, we do not associate entropy to
DE lying inside the BH horizon. In other words, the entropy of DE is
solely determined from its quantity contained between the cosmic and
BH horizons while the entropy of DE inside BH horizon is ignored.
Also note that in a DE filled Universe, we assume that the interior
of BH horizon is always filled with DE as a result of accretion. 

The analysis in later sections is based on the assumption of thermal
equilibrium: the temperature of black hole's event horizon,
cosmological horizon and the phantom energy are the same. But this
assumption in cosmological setting is very ideal, since major
components of the universe including dark matter, dark energy and
radiation (CMB and neutrinos inclusive) have entirely different
temperatures \cite{lima}. But Karami and Ghaffari \cite{kk} recently
demonstrated that the contribution of the heat flow between dark
energy and dark matter for GSL in non-equilibrium thermodynamics is
very small, $O(10^{-7})$. Therefore the equilibrium thermodynamics
is still preserved. Further, if there is any thermal difference in
the fluid and the horizons, the transfer of energy across the
horizons might change the geometry of horizons \cite{log}.

\section{GSL with Bekenstein-Hawking entropy}

\subsection{Use of Event Horizon}

The \textit{future event horizon} is the distance that light travels from present time till infinity is defined as
\begin{equation}
R_E(t)=a(t)\int\limits_t^\infty\frac{dt'}{a(t')}<\infty,
\end{equation}
whose time derivative is
\begin{equation}\label{f11}
\dot R_E=HR_E-1.
\end{equation}
The temperature of future event horizon is proportional to de Sitter's universe horizon
\begin{equation}\label{f7}
T_h=\frac{bH}{2\pi},
\end{equation}
where $b$ is a constant. Depending on the argument chosen from
thermodynamics of phantom energy  \cite{sigu,lima1}, $b$ can be
positive or negative.  Integrating (\ref{f11}) and using (\ref{h}),
we obtain a time evolution of $R_E$:
\begin{equation}\label{f12}
R_E=C_2(t-t_s)^{\frac{2}{3(1+w)}}-\frac{3(t-t_s)(1+w)}{1+3w},
\end{equation}
where $C_2$ is a constant of integration. In the present context,
Sadjadi  \cite{sadjadi} studied the cosmological thermodynamics
using the cosmic future event horizon $R_E$. However in his detailed
analysis, the author ignored a very important first term on right
hand side of (\ref{f12}), which will change significantly the
results for the validity of GSL.

\subsection{Use of Apparent Horizon}

The \textit{apparent horizon} is a null surface with vanishing
expansion  or the boundary surface of anti-trapped region
\cite{peng}. In a spatially flat FRW universe, the apparent horizon
is $R_A=H^{-1}$ (also called Hubble horizon). This horizon is
consistent if we insist on the validity of holography during
inflation i.e. the apparent horizon is the holographic boundary of
the FRW universe. The temperature of apparent horizon is the same as
temperature of a de Sitter's universe horizon \cite{cai}
\begin{equation}\label{f8}
T_h=\frac{H}{2\pi}
\end{equation}
Using (\ref{f16}) the form of GSL at the apparent horizon becomes
\begin{equation}
\dot {S}_{tot}=-32\pi A\dot H M^3 \geq0.
\end{equation}
 In terms of Hubble parameter alone,
\begin{equation}\label{final}
\dot {S}_{tot}= \frac{-32\pi A\dot H}{(C_1+4AH)^3}\geq0.
\end{equation}
In the above case (\ref{final}), mathematically GSL will hold under
two  situations: Case-I: (1) $\dot H\leq0$, $C_1+4AH>0$ or Case-II:
(1) $\dot H\geq0$, (2) $C_1+4AH<0$. However physically, under
phantom dominated era, only case-II is relevant. Thus the apparent
horizon expands in the phantom phase while the future event horizon
contracts.

\section{GSL with Power-Law Entropy Correction}

We extend our previous study here by taking into account the
correction  to horizon's entropy of the power-law form.

\subsection{Use of Event Horizon}

Using the definition of GSL (\ref{f16}) gives
\begin{eqnarray}\label{eh1}
\dot {S}_{tot}&=&2\pi R_E\dot{R}_E \Big[ 1- \frac{\alpha(4\pi)^{\frac{\alpha}{2}-1}}{2r_c^{2-\alpha}}  \Big(\pi R_E^2\Big)^{1-\frac{\alpha}{2}} \Big]
\nonumber\\&&-32\pi A\dot H M^3 \Big[ 1-  \frac{\alpha(4\pi)^{\frac{\alpha}{2}-1}}{2r_c^{2-\alpha}}\Big(\frac{4\pi}{M^2}\Big)^{1-\frac{\alpha}{2}} \Big]\nonumber\\&&+\frac{2\pi\dot H}{bH}\dot R_E\geq0.
\end{eqnarray}
Its easy to interpret the positivity of (\ref{eh1}) by writing the
above equation as a total derivative form
\begin{eqnarray}
\dot {S}_{tot}&=&\frac{d}{dt}\Big[\pi R_{E}^2-\frac{\pi\alpha}{4-\alpha}(2r_c)^{\alpha-2}R_E^{4-\alpha}\\\nonumber&&+ 4\pi M^2 -\frac{8\pi}{2r_c^{2-\alpha}}M^\alpha \Big] \\\nonumber&&+\frac{2\pi\dot H}{bH}\dot R_E\geq0.
\end{eqnarray}
 For phantom $b<0$, and in both cases $\dot{H}<(>0),\dot{R_E}<(>0)$, the general condition to satisfy GSL is
\begin{eqnarray}
&& \frac{d}{dt}\Big[\pi R_{E}^2-\frac{\pi\alpha}{4-\alpha}(2r_c)^{\alpha-2}R_E^{4-\alpha}\\\nonumber&&+ 4\pi M^2 -\frac{8\pi}{2r_c^{2-\alpha}}M^\alpha \Big]  \geq0.
\end{eqnarray}

\subsection{Use of Apparent Horizon}
Using the definition of GSL (\ref{f16}) gives
\begin{eqnarray}
\dot {S}_{tot}&=&2\pi\frac{\dot H}{H^3}\frac{\alpha(4\pi)^{\frac{\alpha}{2}-1}}{2r_c^{2-\alpha}}\Big(\frac{\pi}{H^2}\Big)^{1-\frac{\alpha}{2}}
-32\pi A\dot H M^3\nonumber\\&&\times\Big[ 1-\frac{\alpha(4\pi)^{\frac{\alpha}{2}-1}}{2r_c^{2-\alpha}}\Big(\frac{\pi}{M^2}\Big)^{1-\frac{\alpha}{2}} \Big]\geq0.
\end{eqnarray}
We rewrite it in the following total derivative form
\begin{eqnarray}
\dot {S}_{tot}&=&\frac{d}{dt}\Big[\frac{\pi\alpha}{\alpha-4}\frac{(2r_c)^{\alpha-2}}{H^{4-\alpha}}\nonumber\\&&+4\pi M^2-4\pi(2r_c)^{\alpha-2}M^\alpha\Big]\geq0.
\end{eqnarray}

\section{GSL with Logarithmic Entropy Correction}

We extend our previous study here by taking into account the
correction to horizon's  entropy of the logarithmic form.

\subsection{Case of Event Horizon}
Using the definition of GSL (\ref{f16}) gives
\begin{eqnarray}
\dot {S}_{tot}&=&2\dot{R}_E\Big( \pi R_E+\frac{\beta}{R_E} \Big)-8A\dot H M\Big( \beta-4\pi M^2  \Big)\nonumber\\&&+\frac{2\pi\dot H}{bH}\dot R_E\geq0.
\end{eqnarray}
For phantom $b<0$, and in both cases $\dot{H}<(>0),\dot{R_E}<(>0)$, the general condition to satisfy GSL is
\begin{eqnarray}
\dot {S}_{tot}&=&\frac{d}{dt}\Big[\pi R_E^2+2\beta\log (MR_E)-4\pi M^2\Big]\nonumber\\&&+\frac{2\pi\dot H}{bH}\dot R_E\geq0.
\end{eqnarray}

\subsection{Case of Apparent Horizon}
Using the definition of GSL (\ref{f16}) gives
\begin{equation}
\dot {S}_{tot}=-2\pi\frac{\dot H}{H}\beta M-8A\dot H M^2\Big( \beta-4\pi M^2 \Big)\geq0.
\end{equation}
Writing the above equation as a total derivative form, we get
\begin{equation}
\dot {S}_{tot}=\frac{d}{dt}\Big(\beta\log(\frac{M}{H})-2\pi M^2\Big)+\frac{2\pi\dot H}{bH}\dot R_E\geq0.
\end{equation}

\section{Phantom energy accretion by black hole: GSL constraints}

Pacheco \& Horvath \cite{hor} investigated the generalized second
law of thermodynamics  for a static spherically symmetric black hole
accreting phantom energy. They showed that for a phantom fluid
violating the null energy condition ($\rho+p<0$), the Euler relation
($\rho+p=TS$) and Gibbs relation ($E+pV=TS$, assuming $E=\rho V$ we
get $(\rho+p)V=TS$) allows two different possibilities for the
entropy and temperature of phantom energy: a situation when the
entropy is negative and the temperature is positive or vice versa.
In the former case, if GSL is valid, the accretion of phantom energy
is not allowed while in the later case, there is a critical black
hole mass above which the accretion process is not allowed. In
another study, Lima et al \cite{mu} discussed the thermodynamics of
phantom energy with chemical potential $\mu_0$ and found the EoS
parameter of the form
\begin{equation}
w\geq-1+\frac{\mu_0 n_0}{\rho_0},
\end{equation}
where $n_0$ is the number density and $\rho_0$ is the energy density
of phantom energy.  The authors deduced if $\mu_0<0$ then $w$
describes a phantom fluid. Lima and co-workers \cite{limaco} showed
that the temperature of the phantom fluid without chemical potential
is positive definite but entropy is negative $S_d < 0$. But their
claim is problematic since if considering the usual statistical
definition of entropy, than phantom energy must not exist at all.
Later Pacheco \cite{P} ruled out the previous results of phantom
accretion by black holes with chemical potential \cite{mu,limaco}.
Below, we adapt the procedure of \cite{hor} about constraints
imposed by GSL on the mass of a black hole. We find that there
exists a critical value of black hole mass (or the upper bound)
$M_c$ below which accretion of phantom is allowed. Note that in the
subsequent analysis, we deal phantom energy without chemical
potential on account of \cite{P}. Also the foregoing analysis with
Bekenstein-Hawking entropy has been done in \cite{hor}, therefore,
we will continue our work with logarithmic and power-law entropy
corrections only.

\subsection{Using Logarithmic Entropy}

We consider the new entropy as a sum of black hole entropy and entropy of phantom energy as
\begin{equation}
S_n=f(X)+\kappa \rho^{\frac{1}{1+w}}V,
\end{equation}
where $\kappa$ is a constant and $f(X)$ is a given function of area. We first start with logarithmic entropy
\begin{equation}\label{f}
f(X)=X+\beta\log(X)+\gamma,\ \ X\equiv\frac{A}{4}=4\pi M^2.
\end{equation}
On account of accretion, the change in the entropy of black hole and the phantom energy is
\begin{equation}\label{ds}
\Delta S_n=f'(X)\Delta X+ \frac{\kappa}{1+w}\rho^{\frac{-w}{1+w}}V\Delta\rho.
\end{equation}
 The total energy conservation for this system
is \cite{hor}
\begin{equation}
\Delta M=-\frac{1}{2}(1+w)V\Delta\rho,
\end{equation}
or
\begin{equation}\label{dm}
V\Delta\rho=-\frac{2\Delta M}{1+w}.
\end{equation}
Using (\ref{dm}) in (\ref{ds}), we get
\begin{equation}\label{delta-s}
\Delta S_n=\Big[ 8\pi Mf'(X)-\frac{2\kappa}{(1+w)^2} \rho^{\frac{-w}{1+w}} \Big]\Delta M.
\end{equation}
Since mass of black hole decreases due to accretion of phantom energy i.e. $\Delta M<0$, we require
\begin{equation}
\Delta S_n >0  \ \Rightarrow 4\pi M_{c}f'(X_c)-\frac{\kappa}{(1+w)^2} \rho^{\frac{-w}{1+w}} <0.
\end{equation}
From (\ref{f}), we have $f'(X)=1+\frac{\beta}{X}$. Thus
\begin{equation}\label{quad}
M_c^2-\Big( \frac{\kappa}{4\pi (1+w)^2}\rho^{\frac{-w}{1+w}} \Big)M_c+\frac{\beta}{4\pi}\geq0.
\end{equation}
The discriminant of (\ref{quad}) is
\begin{equation}
\delta=\Big( \frac{\kappa}{8\pi (1+w)^2}\rho^{\frac{-w}{1+w}} \Big)^2-\frac{\beta}{4\pi}\geq0.
\end{equation}
Here two cases are possible: (1) $\delta=0$ gives $(M_c-M)^2<0$
which is nonphysical and mathematically not possible, (2) $\delta>0$
implies $(M_c-M_+)(M_c-M_-)>0$, where $M_\pm$ are the roots of
(\ref{quad}) given by
\begin{equation}
M_\pm=\frac{\kappa}{8\pi(1+w)^2}\rho^{\frac{-w}{1+w}}\pm\sqrt{\Big( \frac{\kappa}{8\pi (1+w)^2}\rho^{\frac{-w}{1+w}} \Big)^2-\frac{\beta}{4\pi}},
\end{equation}
while the critical black hole mass lies in the range
\begin{equation}
M_-<M_c<M_+,
\end{equation}
where $M_+$ and $M_-$ are the corresponding upper and lower bounds on critical mass of black hole.

\subsection{Using Power-Law Corrected Entropy}

The form of power-law corrected entropy is
\begin{equation}
S(X)=X[1-K_\alpha(4X)^{1-\frac{\alpha}{2}}].
\end{equation}
Here condition (\ref{delta-s}) implies
\begin{eqnarray}
&&8\pi M_c\Big[ 1-K_\alpha \Big( 2-\frac{\alpha}{2} \Big)(16\pi)^{1-\frac{\alpha}{2}}M_c^{2-\alpha} \Big]\nonumber\\&&-\frac{2\kappa}{(1+w)^2}\rho^{\frac{-w}{1+w}}<0.
\end{eqnarray}
In terms of $r_c$, the general equation for critical mass to satisfy is
\begin{eqnarray}\label{imp}
4\pi M_c\Big[1-\frac{\alpha}{2}\Big(\frac{2M_c}{r_c}\Big)^{2-\alpha}\Big]-\frac{\kappa}{(1+w)^2}\rho^{\frac{-w}{1+w}} <0. \label{gen}
\end{eqnarray}
To solve (\ref{imp}), we consider some special cases of (\ref{imp}) for different values of $\alpha=1,2,3,4,5$.

\subsection{$\alpha=1$}

We can write (\ref{imp}) in terms of cross-over scale parameter $r_c$ for convenience:
\begin{equation}\label{alpha1}
M_c-\frac{\alpha}{r_c}M_c^2-\frac{\kappa}{4\pi(1+w)^2}\rho^{-\frac{w}{1+w}}<0.
\end{equation}
Here discriminant of the above expression is
\begin{equation}
\delta=1-\frac{\kappa}{\pi r_c(1+w)^2}\rho^{-\frac{w}{1+w}}\geq0.
\end{equation}
The corresponding roots are
\begin{equation}
M_{\pm}=\frac{r_c}{2\alpha}(1\pm\sqrt{\delta}).
\end{equation}
The critical black hole mass lies in the range
\begin{equation}
M_-<M_c<M_+,
\end{equation}
where $M_+$ and $M_-$ are the corresponding upper and lower bounds on critical mass of black hole.

\subsection{$\alpha=2$}
In this case, we get the lower bound on the mass of black hole:
\begin{equation}
M>M_c=\frac{-\kappa}{4\pi (1+w)^2}\rho^{\frac{-w}{1+w}},
\end{equation}
while $\kappa<0$ since mass can not be negative physically.

\subsection{$\alpha=3$}
Here we obtain an upper bound on the mass of black hole as
\begin{equation}
M<M_c=\frac{3}{4}r_c+\frac{\kappa}{4\pi (1+w)^2}\rho^{\frac{-w}{1+w}}.
\end{equation}

\subsection{$\alpha=4$}

The equation to be satisfied by critical mass is
\begin{equation}
M_c^2-\frac{\kappa}{4\pi(1+w)^2}\rho^{\frac{-w}{1+w}}M_c-\frac{r_c^2}{2}<0
\end{equation}
The discriminant of the above equation is
\begin{equation}
\delta=\Big( \frac{\kappa \rho^{\frac{-w}{1+w}}}{4\pi (1+w)^2} \Big)^2+2r_c^2>0,
\end{equation}
while the roots are
\begin{equation}
M_\pm=\frac{\kappa}{8\pi (1+w)^2}\rho^{\frac{-w}{1+w}}\pm\frac{\sqrt{\delta}}{2}.
\end{equation}
Further critical mass satisfies $M_-<M_c< M_+$.

\subsection{$\alpha=5$}

Here we have the following cubic equation
\begin{eqnarray}
M_c^3-A_1 M_c^2-B_1<0,  \label{cubic}
\end{eqnarray}
where
\begin{eqnarray}
A_1=\frac{\kappa}{4\pi(1+w)^2}\rho^{-\frac{w}{1+w}},\ \
B_1=\frac{5}{16}r_c^3
\end{eqnarray}
We define a new variable $y=M+\frac{A_1}{3}$, than the (\ref{cubic}) converts to
\begin{eqnarray}
y^3+py+q<0, \  \ p=-\frac{A_1^2}{3},\  \  q=B+\frac{2}{27}A_1^3
\end{eqnarray}
Since $p<0$, this cubic equation has three distinct solutions which are
\begin{eqnarray}
y_1&=&\sqrt{-\frac{p}{3}}\cos(\psi), \\ y_2&=&\sqrt{-\frac{p}{3}}\cos(\frac{\psi}{3}+\frac{\pi}{3}),\\
y_3&=&\sqrt{-\frac{p}{3}}\cos(\frac{\pi}{3}-\frac{\psi}{3}),
\end{eqnarray}
where
$$
\cos(\psi)=\frac{108B_1+8A_1^3}{8A_1^3}.
$$
Thus there are two possibilities for the value of critical mass
\begin{eqnarray}
M_c<M_{1},\ \ M_{2}<M_c<M_{3},
\end{eqnarray}
where
\begin{eqnarray}
M_{1}&=&\frac{\kappa}{6\pi(1+w)^2}\rho^{-\frac{w}{1+w}}\Big(-\frac{1}{2}+\cos(\psi)\Big),\\
M_{2}&=&\frac{\kappa}{6\pi(1+w)^2}\rho^{-\frac{w}{1+w}}\Big(-\frac{1}{2}+\cos(\frac{\pi}{3}-\frac{\psi}{3})\Big),\\
M_{2}&=&\frac{\kappa}{6\pi(1+w)^2}\rho^{-\frac{w}{1+w}}\Big(-\frac{1}{2}+\cos(\frac{\pi}{3}+\frac{\psi}{3})\Big).
\end{eqnarray}

\section{Conclusion}

In this paper, we studied the generalized second  law of
thermodynamics for a system comprising of a Schwarzschild black
accreting a test non-self-gravitating fluid namely phantom energy in
FRW universe. We are interested if the entropy of this whole system
is positive or not. Since second law of thermodynamics is
fundamental law of physics, its validity needs to be checked for a
thermal system. As is known the form of entropy-area relation for
any causal horizon (either black hole or cosmological) is the same,
we discussed entropies of these horizons with classical relation of
Hawking and with quantum corrections, for the sake of consistency.
However the entropy of phantom energy was calculated using first law
of thermodynamics. Using two different forms of cosmological
horizon, we found general conditions for the validity of GSL in the
present context. Next we used the GSL to impose some restrictions
(upper or lower bounds) on the mass of black hole under which a
black hole can accrete phantom energy. 
In this article, we
are unable to predict whether the observable Universe is dominated
by phantom energy or not. Rather we assumed that it contains phantom
energy as claimed by Caldwell \cite{cald,cald1}. Moreover so far
there are no theoretical constraints on the model parameters (like
$\beta$ and $\gamma$ etc) except the BH mass coming from our model.

\end{document}